\documentclass[aps,prl,amsfonts,superscriptaddress,twocolumn,showpacs,nofootinbib]{revtex4-1}  
\usepackage{graphicx}  
\usepackage{amssymb}   
\usepackage{amsmath}
\usepackage{hyperref}
\usepackage{verbatim}
\usepackage{multirow}
\usepackage{float,xcolor}

\begin{document}
\title{Limits on the Existence of sub-MeV Sterile Neutrinos from the Decay of $^7$Be in Superconducting Quantum Sensors}
\author{S.~Friedrich}\email{friedrich1@llnl.gov}\affiliation{Lawrence Livermore National Laboratory, Livermore, CA 94550, USA}
\author{G.B.~Kim}\affiliation{Lawrence Livermore National Laboratory, Livermore, CA 94550, USA}
\author{C.~Bray}\affiliation{Department of Physics, Colorado School of Mines, Golden, CO 80401, USA}
\author{R.~Cantor}\affiliation{STAR Cryoelectonics LLC, Santa Fe, NM 87508, USA}
\author{J.~Dilling}\affiliation{TRIUMF, Vancouver, BC V6T 2A3, Canada}
\author{S.~Fretwell}\affiliation{Department of Physics, Colorado School of Mines, Golden, CO 80401, USA}
\author{J.A.~Hall}\affiliation{STAR Cryoelectonics LLC, Santa Fe, NM 87508, USA}
\author{A.~Lennarz}\affiliation{TRIUMF, Vancouver, BC V6T 2A3, Canada}\affiliation{Department of Physics and Astronomy, McMaster University, Hamilton, ON L8S 4M1,
Canada}
\author{V.~Lordi}\affiliation{Lawrence Livermore National Laboratory, Livermore, CA 94550, USA}
\author{P.~Machule}\affiliation{TRIUMF, Vancouver, BC V6T 2A3, Canada}
\author{D.~McKeen}\affiliation{TRIUMF, Vancouver, BC V6T 2A3, Canada}
\author{X.~Mougeot}\affiliation{Universit\'e Paris-Saclay, CEA, List, Laboratoire National Henri Becquerel (LNE-LNHB), F-91120, Palaiseau, France}
\author{F.~Ponce}\affiliation{Department of Physics, Stanford University, Stanford, CA 94305, USA}\affiliation{Lawrence Livermore National Laboratory, Livermore, CA 94550, USA}
\author{C.~Ruiz}\affiliation{TRIUMF, Vancouver, BC V6T 2A3, Canada}
\author{A.~Samanta}\affiliation{Lawrence Livermore National Laboratory, Livermore, CA 94550, USA}
\author{W.K.~Warburton}\affiliation{XIA LLC, Hayward, CA 94544, USA}
\author{K.G.~Leach}\email{kleach@mines.edu}\affiliation{Department of Physics, Colorado School of Mines, Golden, CO 80401, USA}
\date{\today}

\begin{abstract}
Sterile neutrinos are natural extensions to the standard model of particle physics and provide a possible portal to the dark sector.  We report a new search for the existence of sub-MeV sterile neutrinos using the decay-momentum reconstruction technique in the decay of $^7$Be.  The experiment measures the total energy of the $^7$Li daughter atom from the electron capture decay of $^7$Be implanted into sensitive superconducting tunnel junction (STJ) quantum sensors. This first experiment presents data from a single STJ operated at a low count rate for a net total of 28 days, and provides exclusion limits on sterile neutrinos in the mass range from 100 to 850 keV that improve upon previous work by up to an order of magnitude.

\end{abstract}

\maketitle

The standard model of particle physics (SM) is one of the crowning achievements in modern science and the cornerstone of current subatomic studies. Despite its success, the SM is known to be incomplete, and physics beyond the standard model (BSM) is required to develop a full description of the Universe~\cite{Lyk10}.  The neutrino sector, in particular, offers an intriguing avenue for BSM physics as the observation of non-zero neutrino masses currently provides the only confirmed violation of the SM as it was originally constructed~\cite{Ramond:1999vh,Bilenky:2014ema}.  Neutrinos are light, neutral leptons, and the only particles in the SM that have an intrinsic chirality, in that they only interact via left-handed (LH) currents of the weak interaction through three eigenstates, $\nu_e$, $\nu_\mu$, and $\nu_\tau$.  Neutrino oscillation experiments over the last 20 years~\cite{Fuk98,Ahm01} have also indicated that these flavor states include (at least) three non-zero mass eigenstates.  The weak-interaction eigenstates and mass eigenstates of the neutrinos are not identical, but are related via a transformation known as the Pontecorvo-Maki-Nakagawa-Sakata (PMNS) matrix~\cite{Pontecorvo:1967fh,Pontecorvo:1957cp,Maki:1962mu}.

Non-zero neutrino masses make extensions to the SM description of leptons unavoidable. Furthermore, it is desirable that such extensions also account for the observed neutrino masses being more than 5 orders of magnitude smaller than that of the electron.  Perhaps the simplest and most studied neutrino mass generation model is the type-I seesaw mechanism~\cite{Minkowski:1977sc,GellMann:1980vs,Glashow:1979nm,Mohapatra:1979ia,Schechter:1980gr}, in which $n$ number of right-handed (RH) chiral flavor states are added, known as sterile neutrinos, that are inactive in the weak interaction.  Although the number of active LH neutrino flavors is known to be three based on strong experimental constraints~\cite{Len13}, the total number of flavor states and mass states are unknown since the active LH neutrinos can mix with an unknown number of sterile neutrinos. These RH neutrinos have a mass that does not depend on the Higgs mechanism, the so-called Majorana mass~\cite{Maj37}, and can therefore exist independently of electroweak symmetry breaking and be on nearly any mass scale.

Although these SM extensions can generate a number of different coupling scenarios for additional neutrino masses~\cite{deG09}, those on the keV scale are perhaps the most intriguing, as they are strong candidates for so-called ``warm'' dark matter and may help to address the origin of the matter/antimatter asymmetry of the Universe~\cite{Adh17,Boy19}.  The most interesting experimental hints currently exist in the $1-10$~keV mass range (see, e.g.,~\cite{deGouvea:2019phk}), and recent work~\cite{Fen20} has also generated interest in searches for mass states of order $\sim100$~keV, which may help shed light on the anomalous excess of events reported in the XENON1T experiment~\cite{Apr20}.  Sterile neutrinos in this mass range generically have lifetimes long enough to be cosmologically relevant, and therefore experiments sensitive to such states offer a complementary avenue to test our cosmological history.

Given the wide range of masses and couplings from model predictions for heavy BSM neutrinos, effective experimental searches for these particles should be model-independent and cover a large area of the allowed parameter space.  One conceptually simple approach is through energy and momentum conservation in nuclear $\beta$ decay~\cite{Cra48}. In these experiments, the momenta of the nuclear daughter recoil and the emitted electron or positron are measured, while the neutrino is not detected and generates missing momentum in the observed spectrum~\cite{Shr80}.  The experimental situation is simplified further in neutron-deficient nuclei with $Q<1022$~keV, where $\beta^+$-decay is energetically forbidden and the parent nucleus can only decay by orbital electron capture (EC)~\cite{Bam77,Fil14}. EC decay provides a two-body final state (rather than three-body for $\beta^\pm$ decay) that consists of the daughter nucleus and the emitted $\nu_e$, both of which are (in principle) mono-energetic.  As a result, the neutrino mass can be directly accessed via precision measurements of the daughter recoil kinetic energy $T_D$, which depends only on the decay energy $Q_{EC}$ and final-state masses $m_\nu$ and $m_D$: 
\begin{equation} \label{T:ERecoil}
    \begin{aligned}
        T_D & = \frac{Q_{EC}^2-m_\nu^2 c^4}{2 \left(Q_{EC}+m_D c^2\right)}.
    \end{aligned}
\end{equation}
This method of decay momentum reconstruction is a simple, model-independent approach to massive neutrino searches, since it relies only on the existence of a heavy neutrino admixture to the active neutrinos - a generic feature of neutrino mass mechanisms - and not on the model-dependent details of their interactions.

The pure EC decaying nucleus $^7$Be is the ideal case for neutrino studies via momentum reconstruction due to its large decay energy $Q_{EC}=861.89(7)$~keV~\cite{AME16}, relatively high maximum value for the recoil kinetic energy $T_{D(max)}=56.826(9)$~eV, and simple atomic and nuclear structure~\cite{Til02}.  Due to the explicit neutrino mass dependence on the recoil kinetic energy in Eqn.~\ref{T:ERecoil}, the existence of a heavier mass state, $m_4$, would cause the nuclear recoils to have a lower $T_D$, thus shifting a fraction of events in the observed spectrum to lower energies, where the relative fraction is governed by the mixing with the $\nu_e$ flavor, $|U_{e4}|^2$.

The EC decay of $^7$Be was first used in the early 1950s as a search for single neutrino emission~\cite{Dav52}, and has been suggested as an ideal case for keV-mass sterile neutrino searches~\cite{Hin98,Smi19}, but has not been explored previously due to technical challenges. In this Letter, we introduce an experiment on Beryllium Electron capture in Superconducting Tunnel junctions (``BeEST"). It uses the decay-momentum reconstruction technique to precisely measure the $^7$Be$\rightarrow^7$Li recoil spectrum of $^7$Be atoms implanted into superconducting quantum sensors, and significantly improves the limits on the existence of sterile neutrinos in the 100-850~keV mass range.

$^7$Be decays by EC primarily to the nuclear ground-state of $^7$Li with a half-life of $T_{1/2}=53.22(6)$~days~\cite{Til02}.  A small branch of 10.44(4)\% results in the population of a short-lived excited nuclear state in $^7$Li ($T_{1/2}=72.8(20)$~fs)~\cite{Til02} that de-excites via emission of a 477.603(2)~keV $\gamma$-ray~\cite{Hel00}.  In the EC process, the electron can be captured either from the $1s$ shell ($K$-capture) or the $2s$ shell ($L$-capture) of Be. For $K$-capture, the binding energy of the $1s$ hole is subsequently liberated by emission of an Auger electron or an X-ray ( where the former is the heavily dominant process~\cite{Hub94}), whose energy adds to the decay signal and separates it from the $L$-capture signal.  Since the nuclear decay and subsequent atomic relaxation occur on short time scales, a direct measurement produces a spectrum with four peaks: two for $K$-capture and two for $L$-capture into the ground state and the excited state of $^7$Li.  Due to the relative spatial overlaps of the $1s$ and $2s$ electron orbitals with the nucleus, the EC process is dominated by $K$-shell capture with a measured $L/K$ capture ratio of 0.040(6) in HgTe~\cite{Voy02} and 0.070(7) in Ta~\cite{Fre20}.

For a high precision measurement of the decay products, the $^7$Be atoms were directly implanted into high-resolution superconducting tunnel junction (STJ) quantum sensors.  STJs are a type of Josephson junction that consists of two superconducting electrodes separated by a thin tunnel barrier, and were initially developed as detectors for astronomy~\cite{Per99} and material science~\cite{Lor03}. In the BeEST experiment, energetic radiation from the $^7$Be decay excites charges above the superconducting energy gap $\Delta$ in proportion to the deposited energy and generates an increased tunneling current signal.  Since the energy gap $\Delta$ is of order 1~meV, STJ detectors have an energy resolution of a few eV full-width at half-maximum (FWHM) in the energy range of interest from 20-120~eV~\cite{Kur82,Pon18}. They can thus distinguish the signals of the different $^7$Be decay channels, including a potential signal from an admixture of heavy sterile neutrinos. The recombination time of excess charges is of order tens of $\mu$s, and STJs can therefore be operated at rates above 1000~counts/s/pixel~\cite{Fra98,Fri20} to capture a high statistics data set from the decay of $^7$Be in realistic run times.

\begin{figure}[t!]
\includegraphics[width=\linewidth]{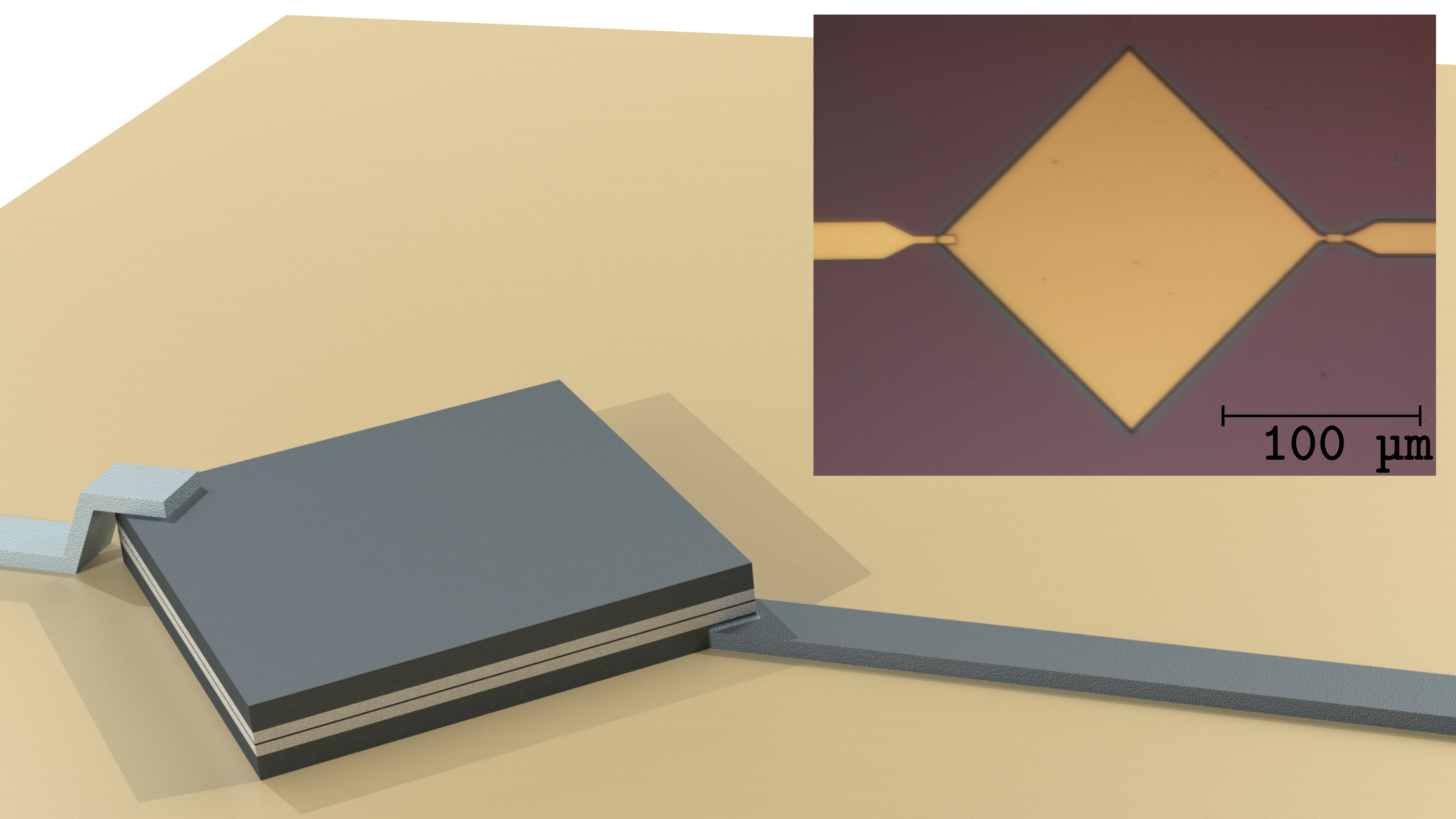}
\caption{\label{ExperimentSchematic}  Schematic of the Ta-Al-Al$_2$O$_3$-Al-Ta STJ used in this work (not to scale), which has a surface area of ($138~\mu$m)$^2$ and a thickness of $0.5~\mu$m.  An image of the sensor is shown in the inset.  The radioactive $^7$Be was implanted into the center of the top Ta layer through a $100~\mu$m  diameter Si collimator.}
\end{figure}

The STJ detector used in this experiment was a five-layer device consisting of Ta (165~nm) - Al (50~nm) - Al$_2$O$_3$ (1~nm) - Al (50~nm) - Ta (265~nm) that was fabricated by photolithography on Si wafers at STAR Cryoelectronics LLC~\cite{Car13} (Fig.~\ref{ExperimentSchematic}) and was characterized in detail prior to implantation~\cite{Fri20}.  The thin film geometry ensured that most of the 478-keV $\gamma$-rays and all neutrinos escape from the detector without interaction, and thus only the recoiling $^7$Li atoms and electrons from the relaxation of its atomic shells were recorded.  The $^7$Be$^+$ ions were implanted into the STJ detectors through Si collimators at TRIUMF's Isotope Separator and ACcelerator (ISAC) facility in Vancouver, Canada~\cite{Dil14} at an energy of 25~keV.  The $^7$Be$^+$ beam was produced using the isotope separation on-line technique~\cite{Blu13} via spallation reactions from a 10~$\mu$A, 480-MeV proton beam incident on a stack of thin uranium carbide targets.  The total number of implanted atoms was intentionally limited to $\sim$10$^8$ to prevent ion-beam damage to the detector surface from the $^7$Li$^+$ ions that were also present in the beam with $\sim$50$\times$ greater intensity. This generated an initial activity in the STJ pixel used for this experiment of roughly 10~Bq.

\begin{figure}[t!]
\includegraphics[width=\linewidth]{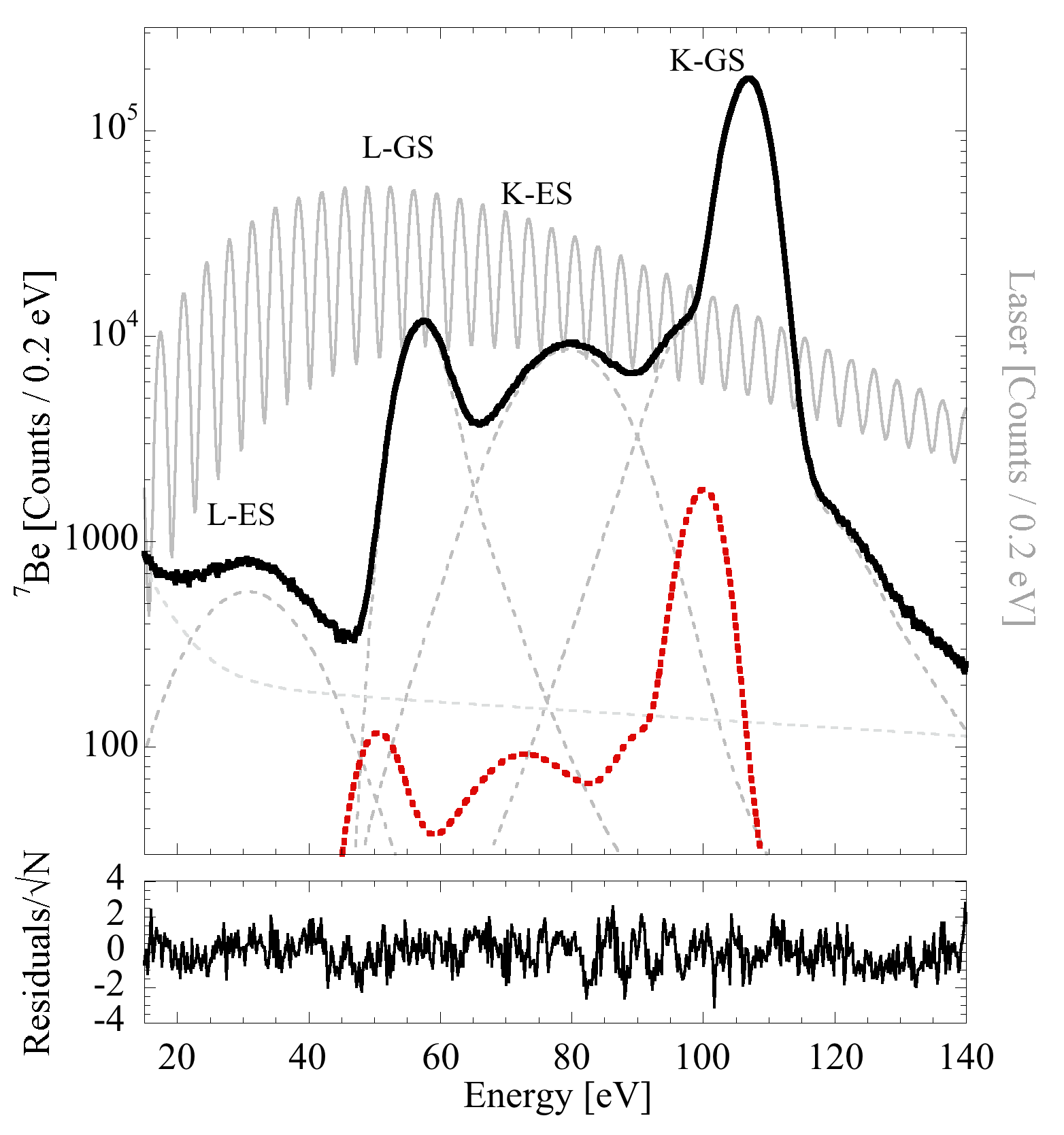}
\caption{\label{7BeSpectrum} The $^7$Li recoil spectrum for 28 days of acquisition (black) shows four peaks for the different $^7$Be decay channels and a broad background due to $\gamma$-interactions in the Si substrate (dashed). The fit residuals (bottom) have a $\chi^2/\nu$ = 1.05. The spectrum generated by a hypothetical 300 keV sterile neutrino signal with $|U_{e4}|^2= 0.01$ is shown in red (dashed), along with the laser calibration signal (grey).}
\end{figure}

The $^7$Li recoil spectrum from the decay of $^7$Be  (Fig.~\ref{7BeSpectrum}) was measured at Lawrence Livermore National Laboratory (LLNL) with the STJ detector at a temperature of $\sim$0.1 K in a two-stage adiabatic demagnetization refrigerator (ADR) with liquid N$_2$ and He precooling. The signals were read out with a specialized current-sensitive preamplifier from XIA LLC~\cite{War15}, processed with an analog spectroscopy amplifier (Ortec 627) with a shaping time of 10~$\mu$s, and captured with a two-channel multi-channel analyzer (Ortec Aspec927). For energy calibration, the STJs were simultaneously exposed to 3.49965(15)~eV photons from a pulsed Nd:YVO$_4$ laser triggered at a rate of 100~Hz~\cite{Pon18}. The laser intensity was adjusted such that multi-photon absorption provided a comb of peaks over the energy range from 20-120~eV. The calibration spectrum was recorded in coincidence with the laser trigger and the $^7$Li recoil spectrum in anti-coincidence. 

Data were acquired for $\sim$20~hours/day over a period of one month. $^7$Be recoil spectra and their corresponding laser calibration were recorded every 30 minutes so that they could be calibrated individually to correct for small drifts in the detector response. For this, the laser signal was fit to a superposition of Gaussian functions, each corresponding to an integer multiple of the single photon energy. Laser peaks below 20~eV and above 120~eV were omitted from the calibration due to poor statistics in the individual 30-minute spectra. The calibrated spectra were re-binned to 0.2~eV and summed (Fig.~\ref{7BeSpectrum}). 

The measured recoil spectrum shows four peaks as expected, one for $K$-capture decay to the nuclear ground state (K-GS), one to the excited state of $^7$Li (K-ES), and two for the corresponding $L$-capture decays (L-GS and L-ES, respectively) (Fig.~\ref{7BeSpectrum}). The two peaks from the excited state decay of $^7$Li were Doppler-broadened from $\gamma$-decay in flight. An exponential background is visible at low energies due to interaction of these 478-keV $\gamma$-rays in the Si substrate below the STJ, which produced high-energy phonons that propagated to the STJ and generated a signal before thermalizing. In addition, the $K$-GS peaks exhibit a low-energy tail due to partial energy loss of the Auger electrons through the detector surface and a high-energy tail due to shake-up and shake-off of $2s$ electrons into empty bound or unbound states, respectively~\cite{Fre20}.

Interestingly, the widths of the K-GS and L-GS peaks are broadened well beyond the width of the laser signals that were captured simultaneously with the same detector, and the lower-energy L-GS peak is slightly wider than the K-GS peak. We investigated possible origins of this excess broadening with density functional theory (DFT) simulations of the Li $1s$ and $2s$ binding energies in the Ta absorber film and their variations with local atomic structural change from the implantation process. We found that up to 2.0~eV of broadening of the K-GS peak could be attributed to an ensemble of 1s binding energies due to structural variations. In particular, the $1s$ levels of interstitial Li shift by 1.2–1.8~eV compared to substitutional Li, with the greatest shift occurring when multiple Li atoms exist in a single interstitial site. Moreover, local disorder of the Ta lattice due to the implant process can contribute additional shifts of 1.5–2.0~eV, with the shifts distributed almost continuously in energy due to the range of local structural variations in a disordered lattice. Finally, the DFT simulations show that the L-GS peak width is affected by hybridization of the Ta $5d$ band with the Li $2s$ levels, which can experience up to 4~eV broadening for interstitial Li and up to 5~eV if the Ta lattice is in addition (locally) disordered.  Thus, the same local atomic variations from the implant process can broaden the L-GS peak more than K-GS peak. These level shifts are close in magnitude to the observed peak widths, although we cannot yet exclude other sources of broadening.

To fit the spectrum in Fig.~\ref{7BeSpectrum}, we approximated the modeled distribution of Li binding energies with a set of three Gaussian functions~\cite{Fre20}. The partial energy loss of the Li KLL Auger electrons at the surface of the STJ was described by an exponentially-modified Gaussian below the K-capture peaks. The atomic L shake-off contributions to the spectrum that result from the sudden change in atomic number were fit using a Levinger function~\cite{Lev53} for consistency with recent precision low-energy decay studies~\cite{Rob20}.  The $\gamma$-ray background was fit to a sum of two exponential decays~\cite{Pon18}. The fit was performed using a least-squares regression in the Python IMINUIT framework and resulted in $\chi^2$/$\nu$=1.05.

\begin{figure}[t!]
\includegraphics[width=\linewidth]{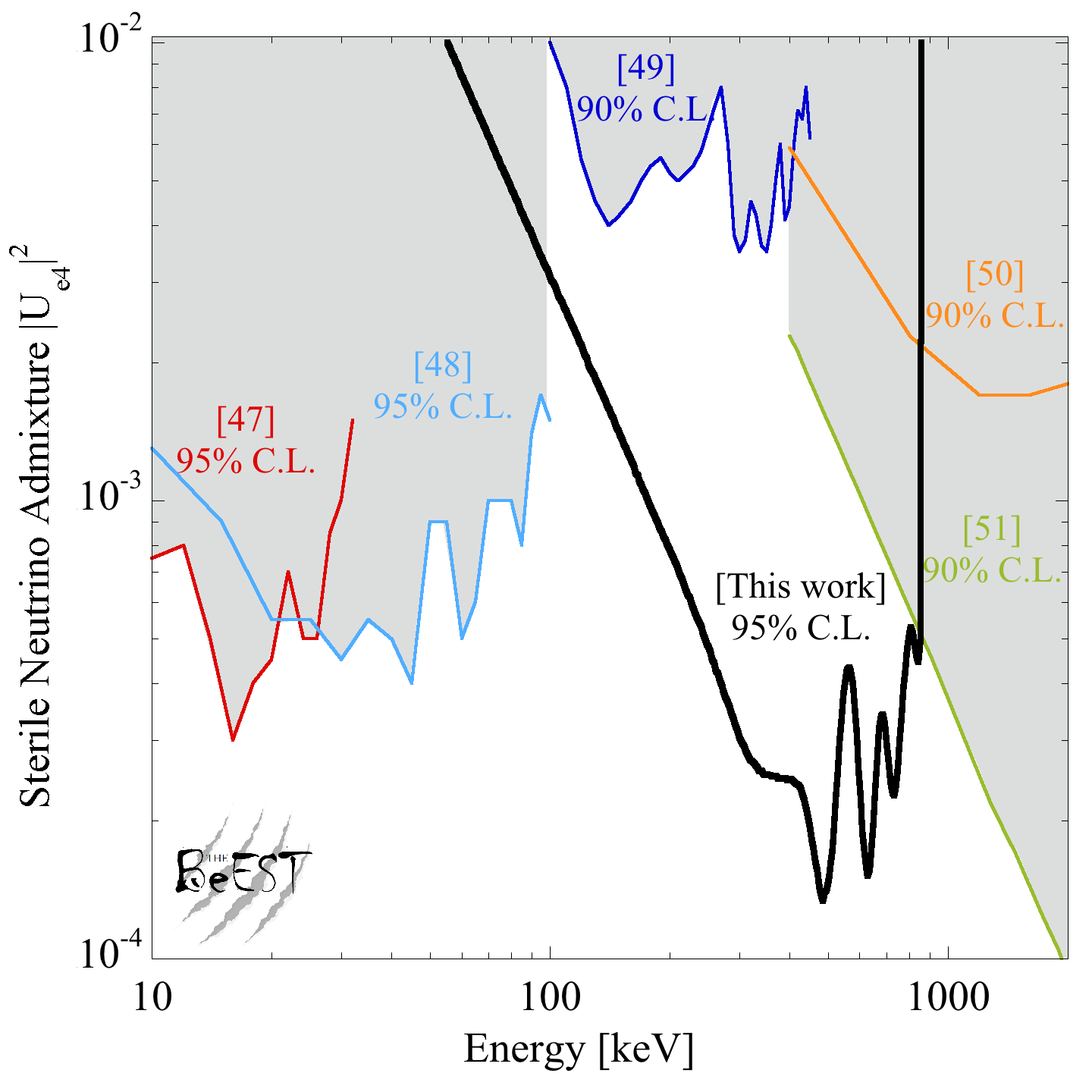}
\caption{\label{Limits} Experimental upper limits for heavy neutrino mixing $|U_{e4}|^2$ as a function of heavy neutrino mass from the BeEST experiment.  The limits are extracted from the data in Fig.~\ref{7BeSpectrum} using a Bayesian approach.  The grey shaded regions are previous best limits in subsections of the mass range from Refs.~\cite{Hol99,Hol00,Sch83,Deu90,Bry19}.}
\end{figure}

Upper limits for $|U_{e4}|^2$, the non-zero mixing between electron neutrino flavor state, $\nu_e$, and a postulated heavy mass state, $m_4$, were subsequently extracted from the measured EC spectrum. For this, we assume that the fit curve describes the null hypothesis of an EC decay involving only the pseudo-degenerate neutrino mass states and represents the background in our search.  The spectrum generated by a heavy neutrino admixture was obtained by shifting the recoil peak centroids according to Eqn.~\ref{T:ERecoil} and scaling the amplitudes of the null hypothesis by $|U_{e4}|^2$. The measured EC spectrum and a hypothetical signal of a sterile $m_4=300$~keV neutrino with $|U_{e4}|^2=10^{-2}$ are shown in Fig.~\ref{7BeSpectrum} as an example. The signal amplitude was then varied systematically for a given sterile neutrino mass, and maximum likelihood values were obtained for each signal amplitude by varying the amplitude and the peak centroids of the background. A Bayesian posterior function was then numerically constructed using this profiled likelihood and a flat prior for positive signal amplitudes, and 95\% upper limits for $|U_{e4}|^2$ were obtained from this posterior function~\cite{feldman1998unified}.  The 95\% confidence limits (C.L.) for exclusion are presented in Fig.~\ref{Limits} with the previous best search limits from nuclear decay data in the mass range of 10-2000~keV.



The new limits on the existence of sterile neutrinos in the 100--850~keV mass range improve upon previous decay measurements by up to an order of magnitude. They make no assumption on theoretical models, neutrino interaction cross-sections, oscillation parameters, or existence of a particular BSM physics scenario.  The measurements only probe heavy neutrino admixtures to the electron neutrino flavor through momentum conservation in nuclear EC decay, and provide the first such limits from the decay of $^7$Be.  These first data from the BeEST experiment were taken with a single STJ pixel counting for roughly a month at a rate that is more than 100 times smaller than its capability.  The statistical precision will be improved in future measurements by scaling the number of STJ detector pixels and increasing the implanted $^7$Be dose, both of which are possible with existing equipment.  For the next phase of the BeEST experiment, increases in statistics will be complemented by detailed imaging and spectroscopy of the implanted ions and the Ta matrix of our STJ detectors and supported by quantum simulations to relate the measurements to the distribution of electronic states that can generate fine structure in the STJ response.  Improvements in these areas, as well as new superconducting materials, will ultimately allow the BeEST experiment to probe increasingly smaller couplings for neutrinos in the 5-860~keV mass range.

\begin{acknowledgements}
This work was supported by the LLNL Laboratory Directed Research and Development program through grants No. 19-FS-027 and No. 20-LW-006, the Office of Nuclear Physics in the U.S. Department of Energy's Office of Science under grants No. DE-SC0017649 and No. DE-SC0021245, and the Natural Sciences and Engineering Research Council of Canada.  TRIUMF receives federal funding via a contribution agreement with the National Research Council of Canada. This research was performed under appointment to the Nuclear Nonproliferation International Safeguards Fellowship Program sponsored by the Department of Energy, National Nuclear Security Administration’s Office of International Nuclear Safeguards (NA-241).  This work was performed under the auspices of the U.S. Department of Energy by Lawrence Livermore National Laboratory under Contract No. DE-AC52-07NA27344.  We thank F. Ames, L. Clark, U. Greife, P. Kunz, J. Lassen, and B. Schultz for their efforts in facilitating the ion-beam implantation, and  J. Behr, A. Bernstein, G. Chapline, T. Kibedi, A. Ray, and F. Sarazin for useful discussions.
\end{acknowledgements}

\bibliography{references}
 
\end{document}